\def\~{{$\tilde{\phantom{a}}$}}

\documentclass [12pt] {article}
\usepackage{epsfig}
\usepackage{color}

\def\thebibliography#1{\section{References}\markboth
 {REFERENCES}{REFERENCES}\list
 {[\arabic{enumi}]}{\settowidth\labelwidth{[#1]}\leftmargin\labelwidth
 \advance\leftmargin\labelsep
 \usecounter{enumi}}
 \def\newblock{\hskip .11em plus .33em minus -.07em}
 \sloppy
 \sfcode`\.=1000\relax}
\def\upcite#1{\raise6pt\hbox{\scriptsize
\cite{#1}}}
\def\lsim{\mathrel {\vcenter {\baselineskip 0pt \kern 0pt
    \hbox{$<$} \kern 0pt \hbox{$\sim$} }}}
\def\gsim{\mathrel {\vcenter {\baselineskip 0pt \kern 0pt
    \hbox{$>$} \kern 0pt \hbox{$\sim$} }}}
\def\gtlt{\mathrel {\vcenter {\baselineskip 0pt \kern 0pt
    \hbox{$>$} \kern 0pt \hbox{$<$} }}}

\def\hline{\noalign{\hrule \vskip2pt}}

\def\|{\ifmmode\Vert\else \char`\|\fi}
\ifx\oldzeta\undefined                          
  \let\oldzeta=\zeta                            
  \def\zzeta{{\raise 2pt\hbox{$\oldzeta$}}}     
  \let\zeta=\zzeta                              
\fi

\ifx\oldchi\undefined                           
  \let\oldchi=\chi                              
  \def\cchi{{\raise 2pt\hbox{$\oldchi$}}}       
  \let\chi=\cchi                                
\fi

\def\frac#1#2{{#1 \over #2}}

\def\half{\ifinner {\scriptstyle {1 \over 2}}
   \else {1 \over 2} \fi}

\def\simge{\mathrel{%
   \rlap{\raise 0.511ex \hbox{$>$}}{\lower 0.511ex \hbox{$\sim$}}}}
\def\simle{\mathrel{
   \rlap{\raise 0.511ex \hbox{$<$}}{\lower 0.511ex \hbox{$\sim$}}}}

\def\buildchar#1#2#3{{\null\!                   
   \mathop#1\limits^{#2}_{#3}                   
   \!\null}}                                    
\def\overcirc#1{\buildchar{#1}{\circ}{}}

\def\slashchar#1{\setbox0=\hbox{$#1$}           
   \dimen0=\wd0                                 
   \setbox1=\hbox{/} \dimen1=\wd1               
   \ifdim\dimen0>\dimen1                        
      \rlap{\hbox to \dimen0{\hfil/\hfil}}      
      #1                                        
   \else                                        
      \rlap{\hbox to \dimen1{\hfil$#1$\hfil}}   
      /                                         
   \fi}                                         %

\def\subrightarrow#1{
  \setbox0=\hbox{
    $\displaystyle\mathop{}
    \limits_{#1}$}
  \dimen0=\wd0
  \advance \dimen0 by .5em
  \mathrel{
    \mathop{\hbox to \dimen0{\rightarrowfill}}
       \limits_{#1}}}                           

\def\overlay#1#2{\ifmmode%
\setbox0=\hbox{$#1$}%
\setbox1=\hbox to\wd0{\hss$#2$\hss}\else%
\setbox0=\hbox{#1}%
\setbox1=\hbox to\wd0{\hss#2\hss}\fi%
#1\hskip-\wd0\box1 }

\def\pmb#1{\leavevmode\setbox0=\hbox{#1}%
\kern-.02em\copy0\kern-\wd0
\kern.04em\copy0\kern-\wd0
\kern-.02em\raise.04em\box0 }

\def\vereq#1#2{\lower3pt\vbox{\baselineskip1.5pt \lineskip1.5pt
\ialign{$\m@th#1\hfill##\hfil$\crcr#2\crcr\sim\crcr}}}

\def\tensor#1{\protect\@ontopof{#1}{\leftrightarrow}{1.15}\mathord{\box2}}
\def\overstar#1{\protect\@ontopof{#1}{\ast}{1.15}\mathord{\box2}}
\def\overdots#1{\protect\@ontopof{#1}{\cdots}{1.0}\mathord{\box2}}
\def\overcirc#1{\protect\@ontopof{#1}{\circ}{1.2}\mathord{\box2}}
\def\loarrow#1{\protect\@ontopof{#1}{\leftarrow}{1.15}\mathord{\box2}}
\def\roarrow#1{\protect\@ontopof{#1}{\rightarrow}{1.15}\mathord{\box2}}

\def\@ontopof#1#2#3{%
{\mathchoice
{\@@ontopof{#1}{#2}{#3}\displaystyle\scriptstyle}%
{\@@ontopof{#1}{#2}{#3}\textstyle\scriptstyle}%
{\@@ontopof{#1}{#2}{#3}\scriptstyle\scriptscriptstyle}%
{\@@ontopof{#1}{#2}{#3}\scriptscriptstyle\scriptscriptstyle}%
}%
}

\def\@@ontopof#1#2#3#4#5{%
\setbox0=\hbox{$#4#1$}%
\setbox1=\hbox{$#5#2$}%
\setbox2=\hbox{}\ht2=\ht0 \dp2=\dp0 %
\ifdim\wd0>\wd1 %
\setbox1=\hbox to\wd0{\hss\box1\hss}%
\mathord{\rlap{\raise#3\ht0\box1}\box0}%
\else   %
\setbox1=\hbox to.9\wd1{\hss\box1\hss}%
\setbox0=\hbox to\wd1{\hss$#4\relax#1$\hss}%
\mathord{\rlap{\copy0}\raise#3\ht0\box1}%
\fi
}%

\def\lambdabar{\protect\@lambdabar}
\def\@lambdabar{%
\relax
\bgroup
\def\@tempa{\hbox{\raise.73\ht0
\hbox to0pt{\kern.25\wd0\vrule width.5\wd0
height.1pt depth.1pt\hss}\box0}}%
\mathchoice{\setbox0\hbox{$\displaystyle\lambda$}\@tempa}%
{\setbox0\hbox{$\textstyle\lambda$}\@tempa}%
{\setbox0\hbox{$\scriptstyle\lambda$}\@tempa}%
{\setbox0\hbox{$\scriptscriptstyle\lambda$}\@tempa}%
\egroup
}

\def\corresponds{{\lower.2ex\hbox{=}}{\rm\kern-.75em^\triangle}}
\def\succsim{\succ\kern-.9em_\sim\kern.3em}
\def\precsim{\prec\kern-1em_\sim\kern.3em}
\def\slantfrac#1#2{\kern1em^{#1}\kern-.3em/\kern-.1em_{#2}}

\begin{document}
                                                                
\begin{center}
{\Large\bf Uncertainties in the Measurement of the Momentum 
\\

\medskip
and Position  of an Electron}
\\

\medskip

Kirk T.~McDonald
\\
{\sl Joseph Henry Laboratories, Princeton University, Princeton, NJ 08544}
\\
(September 20, 2001)
\end{center}

\section{Problem}

An electron is emitted from a source at the origin at time $t = 0$,
 and later is observed at time $t$ by a detector at position $x$.
We can say that the velocity of the electron when it was detected was
$v = x / t$, and therefore its momentum is $P = m v = m x / t$.  By 
making $t$ arbitrarily large, perhaps we can make the uncertainty in momentum $P$
arbitrarily small.

Consider the detector to be a single atom that ``signals" the detection
of the electron by emission of a photon upon de-excitation of the atom
after its interaction with the electron.  Show that the resulting uncertainties
in the measurement of position and momentum obey
\begin{equation}
\Delta x \Delta P \ge \hbar.
\label{p1}
\end{equation}

\section{Solution}

This problem was suggested by a colloquium by Freeman Dyson at Princeton University on
Sept.\ 20, 2001, in which he appeared to claim that the procedure stated above
could evade the Heisenberg uncertainty principle (\ref{p1}).

\bigskip

The uncertainty in the electron's position after detection by the atom is at
least the wavelength $\lambdabar$ of the photon emitted by the excitation
of the atom:
\begin{equation}
\Delta x \ge \lambdabar = {\hbar c \over \Delta E},
\label{s1}
\end{equation}
where $\Delta E$ is the energy of that photon.

For that atom to have emitted the photon as a result of its interaction
with the electron, the electron must have imparted at least energy $\Delta E$
to the atom, and in general it imparted a larger energy than this to the atom.
Therefore, the energy of the electron is uncertain by an amount
of order $\Delta E$.  This implies that the momentum of the electron is
uncertain by the amount
\begin{equation}
\Delta P = {\Delta E \over v}
\ge {\Delta E \over c}\, .
\label{s2}
\end{equation}

The product of the uncertainties (\ref{s1}) and (\ref{s2}) in position
and momentum of the electron after its detection by the atom is
therefore
\begin{equation}
\Delta x \Delta P \ge {\hbar c \over \Delta E} {\Delta E \over c} = \hbar.
\label{s3}
\end{equation}

The size of the detector atom does not enter directly into the above argument.
Hence, the limit (\ref{s3}) cannot be evaded by imagining larger or smaller
atoms.  Rather, the limit arises simply because the mechanism of detection
is an electromagnetic interaction that involves the emission of a photon as
the signal.

\end{document}